\documentclass[12pt]{iopart}
\usepackage{mathptmx}
\usepackage{graphicx}

\begin{document}
\title{Non-universal power laws in transport properties of one-dimensional quantum dots}

\author{T Kleimann\dag, J Stockburger\ddag, M Sassetti\dag, and B Kramer\S}

\address{\dag Dipt. di Fisica, INFM, Universit\`{a} di Genova, Via Dodecaneso 33, 16146 Genova, Italy\\
\ddag II. Inst. f\"ur Theor. Physik, Univ. Stuttgart, Pfaffenwaldring 57, 70550 Stuttgart, Germany\\
\S I. Inst. f\"ur Theor. Physik, Univ. Hamburg, Jungiusstr. 9, 20355 Hamburg, Germany}
\begin{abstract}
  We investigate discrepancies between recent experimental results on
  transport through one-dimensional quantum dots and universal power
  laws predicted by an idealized Luttinger Liquid description. The
  temperature dependence of Coulomb blockade peaks in one-dimensional
  quantum dots obeys non-universal power-laws from which different
  values of the interaction strength can be deduced. We find that,
  depending on the temperature range, measurements probe local or
  global properties of the interaction.  In particular, we investigate
  the role of contacting semiconductor quantum wires and nanotubes
  connected to leads through tunnel junctions and compare to recent
  experiments. We conclude that a conventional Luttinger Liquid
  description of the quantum wire does explain the observed behaviour
  if specific properties of either experimental setup are carefully
  taken into account.
\end{abstract}
\section{Introduction}
Rapid advances have been made in the fabrication of one-dimensional
electronic nanostructures in recent years. In one dimension the usual
Fermi liquid picture of quasiparticles is not applicable anymore.
Instead, the state is described by collective charge density
fluctuations in terms of the Tomonaga--Luttinger liquid model (TLL).
The electronic interactions are then manifest in the typical power-law
exponent $g$. In particular one-dimensional quantum dots are ideal
systems to examine this TLL parameter. One can estimate the $g$ from
the charging energy, and also extract the interaction parameter from
the temperature dependence of the conductance peaks in the quantum
Coulomb blockade (CB) regime. However, in recent experiments probing
the transport through quantum dots in semiconductor quantum wires as
well as in nanotubes contradictory findings for the interaction
parameter where reported \cite{aus,kleim,postma2001}.

When the electron density in the GaAs/AlGaAs-quantum wires fabricated
by using the cleaved-edge-overgrowth technique (CEO) is decreased by
applying a voltage to an external gate, eventually even the lowest
electronic subband can be depopulated \cite{aus}. Here, the mean
electron density is so low that only very few maxima of the random
potential of the impurities are higher than the Fermi level. A
one-dimensional quantum island can be formed between two potential
maxima in such a wire. At temperatures lower than the charging energy,
the linear conductance shows discrete peaks that correspond to
transferring exactly one electron through the quantum island and hence
a one-dimensional single electron transistor (SET) is created.  In
this regime, it has been detected that the temperature dependence of
the intrinsic width of several conductance peaks (the area below the
peaks) is modified by the correlations between the electrons and shows
a power-law behaviour in temperature.

Carbon nanotubes are cylindrical graphene sheets with a diameter of
only a few nanometres and a length of several microns. In the radial
direction electrons are confined by the monolayer thickness of the
graphene sheet. The unique electronic properties of these
nanostructures are due to the quantum confinement of electrons normal
to the tube axis. The electrons can only propagate along the nanotube
axis and form a truly one-dimensional system. Tubes can be metallic or
semiconducting depending on the actual molecular configuration.  SET's
were made by placing a metallic nanotube between two metal electrodes
\cite{tans}. Then, the nanotube itself acts as the island of the SET
and the contact resistances form the tunnel junctions of the
transistor.  Recently, an SET operated at room temperature was
fabricated by putting a metallic nanotube between Au contacts and
manipulating the tube with an atomic force microscope
\cite{postma2001}.  In this way, two buckles in a distance of 25 nm
were created. Buckles in a nanotube behave much like electronic tunnel
junctions \cite{postma2000,bockrath291}, an hence a quantum dot is
formed between the two. However, the interaction parameter which was
extracted from the temperature dependence of the CB peaks did not
correspond to the theoretically expected one.

We suggest two mechanisms, each appropriate for a quantum dot immersed
in a CEO quantum wire, and nanotube, respectively.  To explain the
discrepancy between expected and measured power-law exponents we focus
on the nature of the contacts which connect the quantum wire to the
leads.
\section{Luttinger liquid with impurities}
Using the bosonisation method the excitations of one-dimensional interacting
electron system are described by density waves in terms of the conjugate
fields \cite{voit}, $[\vartheta(x),\Pi(x')]=i \delta(x-x')$. In the simplest
case, one band of spinless electrons, the Hamiltonian is
\begin{equation}
H_0=\frac{v_{\rm F}}{2} 
\int {\rm d}x \left( \Pi^2(x) + 
\frac{1}{g_0^2}[\partial_x \vartheta(x)]^2 \right),
\end{equation}
where $v_{\rm F}$ is the Fermi velocity. We set $\hbar=1=k_{\rm B}$
throughout this paper. The fields are related to the electron density
$\rho$ via $\partial_x \vartheta(x) = \sqrt{\pi} \left[ \rho(x)
-\rho_0\right]$ where $\rho_0=k_{\rm F}/\pi$ is the mean density.  The
parameter $g_0=[1+V_0/\pi v_{\rm F}]^{-1/2}$ is the interaction
constant that arises from the Fourier transform of any one-dimensional
interaction potential in the limit of vanishing screening
length. $0<g_0<1$ corresponds to repulsive interactions, $g_0=1$ no
interaction, and $g_0>1$ attractive.  A single localized impurity at
$x_{\rm b}$ contributes a periodic potential term
\begin{equation}
H_{\rm b} = U_{\rm b} \cos \left[ 2 k_{\rm F}x_{\rm b} + 2 \sqrt{\pi}\vartheta(x_{\rm b}) \right].
\end{equation}
The Hamiltonian $H_0 + H_{\rm b}$ can be interpreted as a potential
model with variable $\vartheta(x_{\rm b})$, coupled to a harmonic
field described by $H_0$. At low energies, traversal of the potential
barriers is by tunneling, corresponding to the transport of electrons
through a tunnel junction \cite{kane}. This electronic transport is
characterized by the evolution of $\vartheta(x_{\rm b})$,
influenced by the bulk modes away from the barrier position. Deriving
an effective action \cite{kleim} for $\vartheta(x_{\rm b})$ the
forward tunneling rate through the junction can be written in the
limit of high barriers \cite{weiss} as
\begin{eqnarray}
  \gamma(V)= \left(\frac{\Delta}{2}\right)^2 
\int_{-\infty}^{\infty} {\rm d}t \; \exp \left[iVt -\int_{-\infty}^\infty {\rm d}\omega \frac{J(\omega)}{\omega^2} \frac{1-e^{-i\omega t}}{1-e^{\omega/T}} e^{-\omega/\omega_{\rm c}} \right].
\label{gamma}
\end{eqnarray}
Here, the tunneling amplitude $\Delta$ is related to $U_{\rm b}$ via
the WKB-method, $V$ the associate energy of the tunneling event (e.g.\ 
voltage across the junction), and $\omega_{\rm c}$ denotes a
plasmon bandwidth cutoff. The spectral density $J(\omega)$ contains
the information due to the plasmon excitations and the electronic
interactions in the system.  It depends in general on the retarded
Green's function $G(x,x';t,t')=-i \Theta(t-t')\langle \left[
  \vartheta(x,t),\vartheta(x',t')\right] \rangle_{H_0}$. In the case
of a single barrier we find
\begin{equation}
J_{\rm b}(\omega,x_{\rm b})= -{\rm Im} \frac{1}{G(\omega;x_{\rm b},x_{\rm b})}.
\label{jb}
\end{equation}
Here, the spectral density simply becomes $J(\omega)=2\omega/g_0$.
Using the detailed balance relation for the rate $\gamma(-V)=
\exp(-V/k_{\rm B}T)\gamma(V)$ we obtain the current $I(V)$ through the
junction from the difference between forward an backward tunneling
rates. Then the linear conductance $\lim_{V\rightarrow 0} I(V)/V$ of
the tunnel junction reads
\begin{equation}
  {\cal G}_{\rm b}(T)=\frac{1}{R_\Delta}\frac{\Gamma^2(1/g_0)}{\Gamma(2/g_0)} \left(
    \frac{2\pi T}{\omega_{\rm c}} \right)^{2/g_0-2}.
\label{sbcond}
\end{equation}
Here, $R_\Delta=2 \omega_{\rm c}^2/\pi e^2\Delta^2$ and $\Gamma$ is the
gamma function. The conductance shows a typical power-law in
temperature dependence explicitely on the interaction through $g_0$.
If electrons tunnel from a metallic lead ($g_0=$1) into the end of a
TLL one needs to replace $2/g_0 \rightarrow 1+1/g_0$ in equation
(\ref{sbcond}), and the power-law for the conductance of such a
contact becomes ${\cal G}_{\rm c}(T)\propto T^{1/g_0 -1}$\cite{kane}.

A quantum dot is created by two such
impurities located at $x_{\rm d}^{\pm}=x_{\rm d}\pm a/2$, whose
barrier contributions to the Hamiltonian can be combined to form the
term
\begin{equation}
 H_{\rm d}= U_{\rm d}\cos(\pi N_{+}) \cos[\pi(n_0+N_{-})],
\end{equation}
where $N_{\pm}=[\vartheta(x_{\rm d}^{+})\pm\vartheta(x_{\rm
d}^{-})]/\sqrt{\pi}$.  Changes of the quantities $N_{+}$ and $N_{-}$
are associated with the transfer of particles between left and right
leads, and the fluctuations of the particle number in the dot,
respectively. One can write the spectral density extracted from the
effective theory for $N_\pm$ in terms of $J_{\rm b}$ in
(\ref{jb})\cite{alepl}
\begin{equation}
J_{\rm d}(\omega)=\frac{1}{2} J_{\rm b}(\omega, x_{\rm d}) \left[ 1 + \epsilon \sum_{n=1}^\infty \delta(\omega - n\epsilon) \right].
\label{jdot}
\end{equation} 
The energy $\epsilon = \pi v_{\rm F}/a g_0=2g_0 E_{\rm c}$ is the
discrete level spacing of the plasmon states in the quantum dot and
$E_{\rm c}$ the charging energy. In the limit of linear transport the
chemical potentials in the left and right leads and the dot are
aligned. Then Coulomb blockade is relaxed and the conductance versus
the gate voltage shows a peak. For sequential tunneling we can use the
master equation method \cite{grade} for calculating the conductance
for $T \ll \epsilon, E_{\rm c}$ and obtain
\begin{equation}
  \label{cbcond}
  {\cal G}_{\rm d}(\mu,T)= \frac{e^2}{4T}
\frac{e^{-\mu/2T}}{\cosh \mu/2{T}} \gamma(\mu).
\end{equation}
Here, $\mu$ is the distance from the resonance energy and $\gamma$ is
defined in (\ref{gamma}) with the appropriate $J_{\rm d}(\omega)$ and
$\Delta$ for $U_{\rm d}$. The conductance of the CB peak reads
\begin{equation}
{\cal G}_{\rm d}(T,\mu) = \frac{1}{4R_\Delta} \frac{|\Gamma(1/2g_0 + i \mu/2\pi T)|^2}{\cosh \mu/2T} \left(\frac{\epsilon}{\omega_{\rm c}} \right)^{1/g_0} \left(
    \frac{2\pi  T}{\omega_{\rm c}} \right)^{1/g_0-2} \frac{e^{-|\mu|/ \omega_{\rm c}}}{\Gamma(1/g_0 )}.
\label{dotcond}
\end{equation}
From (\ref{dotcond}) the maximum of the CB peak ($\mu=0$) scales as ${\cal
  G}_{\rm d}^{\rm max}\propto T^{1/g_0-2}$ in temperature.
\section{CEO wires and SET -- inhomogeneous interaction}
\begin{figure}[tb]

\begin{center}
\includegraphics[width=3cm,keepaspectratio]{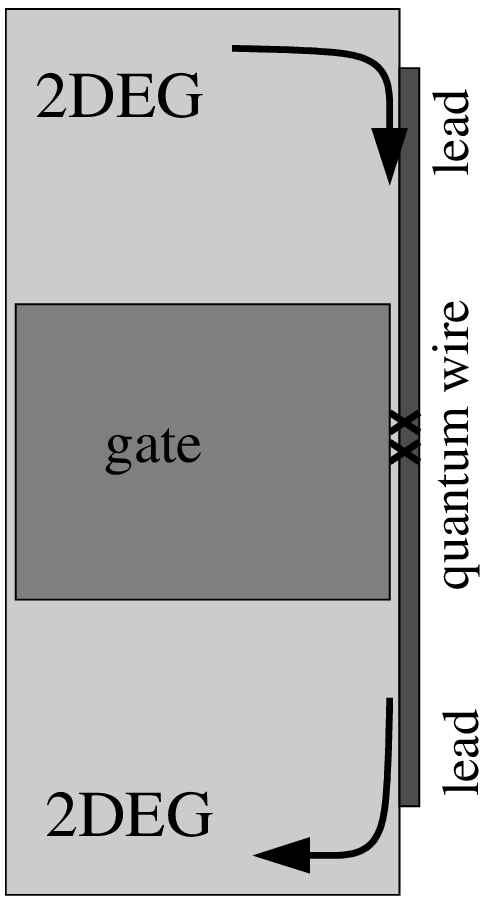}
\includegraphics[width=11.5cm,keepaspectratio]{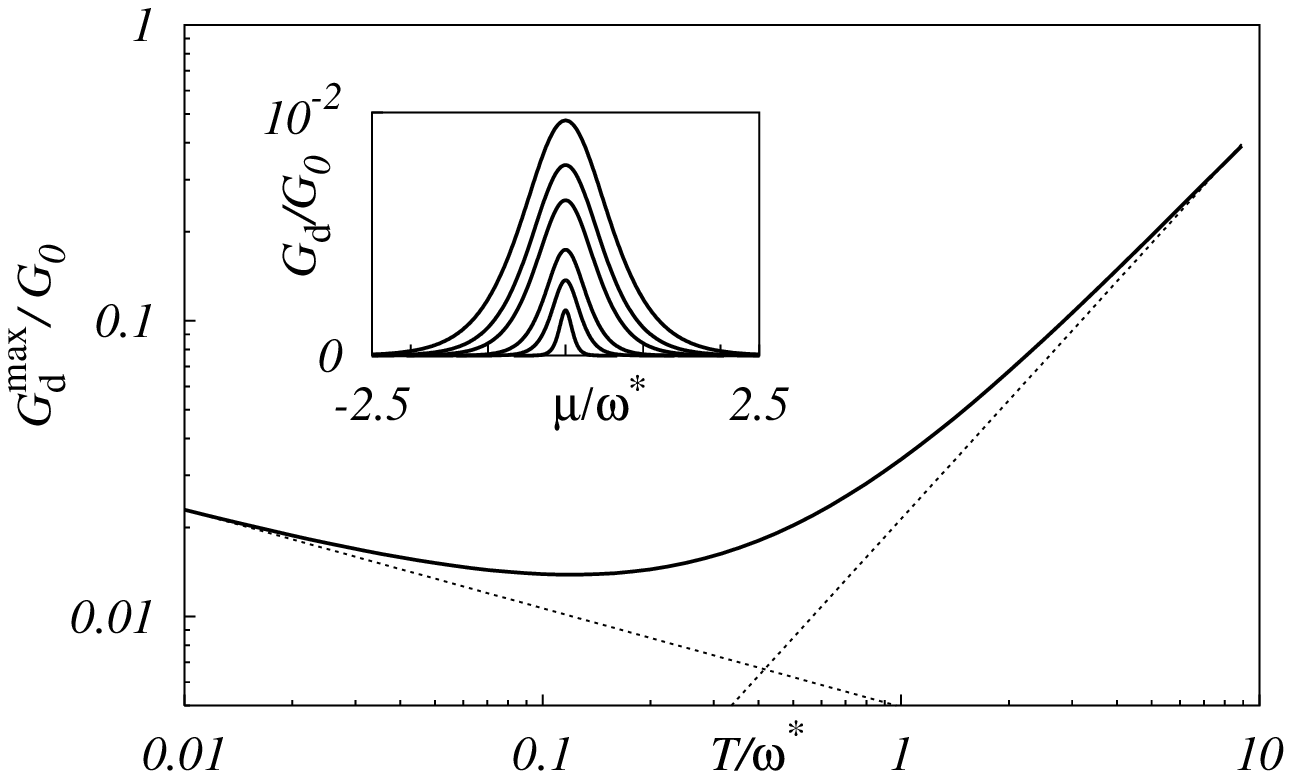}
\end{center}
\caption{
  Left: sketch of the CEO arrangement, electrons travel from the 2DEG
  into the quantum wire situated along the edge. Inset: CB peaks at
  temperatures $T/\omega^\ast=$2.8, 2.3, 1.9, 1.3, 0.9, 0.4 (from top to
  bottom). $\mu$ denotes the distance from the resonance energy.  Both
  are given in units of the characteristic energy induced by the
  inhomogeneity in the interaction strength $\omega^\ast=
  v_{\rm F}/g_0 L$.  Main: temperature dependence of a conductance
  peak for a one-dimensional quantum dot in a TLL with inhomogeneous
  interaction strength. The conductance is given in units
  $G_0=e^2(\Delta/4\omega_{\rm c})^2(\epsilon/ \omega_{\rm
    c})^{1/g_{\rm d}}$ with the tunneling matrix element
  $\Delta$ and the high-energy cutoff $\omega_{\rm c}$ (here:
  $\omega^\ast/\omega_{\rm c}=10^{-3}$). The conductance maximum obeys
  a non-universal power-law \mbox{$G^{\rm max} \propto T^{1/g_{\rm
        eff}-2}$} with a crossover from $g_{\rm eff}=g_0=0.6$
  (interaction strength in the leads) at low temperatures to $g_{\rm
    eff}=g_{\rm d}=0.3$ (local strength at the dot position) at high
  temperatures. The asymptotic power-laws are plotted with dashed
  lines.\label{edin1}}
\end{figure}

In the CEO wires the entire one-dimensional system is situated along the edge
of the sample. Electrons travel from the "leads" region below the
two-dimensional electron gas (2DEG) into the wire region where the electronic
density is kept extremely low and the Coulomb interaction is very strong (see
left sketch in figure \ref{edin1}). We characterize a larger wire region
containing the quantum dot by a spatially varying, short-ranged interaction,
\begin{equation}
  \label{eq:1}
V(x,y)= \left[ V_0 +\varphi(x) \right] \delta (x-y).  
\end{equation}
$\varphi(x)$ is assumed to be smooth, with a maximum near $x=0$ and a
characteristic length $L^{*}$, and $\varphi(x)\to 0$ when $|x|\to\infty$.  The
inhomogeneity also introduces a characteristic frequency
$\omega^\ast=v_{\rm F}/g_0 L^\ast$. With (\ref{eq:1}) the interaction
parameter $g$ varies as a function of $x$,
\begin{equation}
  g(x)  = g_0 \left[1
    + {\varphi(x) \over V_0 + \pi v_{\rm F}} \right]^{-1/2}.
\end{equation}
To calculate the spectral density for the dot we need the Green's
function which obeys the equation of motion
\begin{equation}
  \left[ \frac{\omega^2}{v_{\rm F}}+ \frac{\partial}{\partial x} \frac{v_{\rm F}}{g^2(x)}\frac{\partial}{\partial x} \right] G(\omega;x,x')=-\delta(x-x')
\label{eom}
\end{equation}
with outgoing-wave boundary conditions.
One can immediately solve (\ref{eom}) for the asymptotic cases of short or
long plasmon wavelength, $\omega\gg\omega^\ast$ and $\omega\ll\omega^\ast$.
In the first case, a WKB-like solution yields
\begin{equation}
G(\omega;x,x')=\frac{i}{2} \frac{\sqrt{g(x)g(x')}}{\omega} \exp\left(
\frac{i\omega}{v_{\rm F}}\left|\int_{x'}^x \! {\rm d}y \, g(y) \right|\right),
\end{equation}
and in the low-frequency case we obtain
\begin{equation}
G(\omega;x,x')=\frac{i}{2} \frac{g_0}{\omega} \exp
\left(\frac{i\omega}{v_{\rm F} g_0} \left|\int_{x'}^x \! {\rm d}y\, g^2(y) \right|
\right).\label{lowg}
\end{equation}
This defines the asymptotic behaviour of the corresponding
spectral density for the single barrier, which we later generalize for
the case of a quantum dot with extension $a\ll L^\ast$,
\begin{equation}
 J_{\rm b}(\omega)=\left\{ \begin{array}{l} \frac{2\omega}{g_0}\quad \mbox{for} \quad  \omega \ll \omega^\ast \\   \frac{2\omega}{g_{\rm d}} \quad \mbox{for} \quad  \omega \gg \omega_\ast \end{array} \right..
\end{equation}
Here $g_{\rm d}$ is the local value of the interaction parameter $g(x)$ at the
position of the barrier or dot.
The charging energy of a small dot can easily be deduced\cite{kleim}
from (\ref{lowg}),
\begin{equation}
E_{\rm c} = \frac{\pi v_{\rm F}}{2 a g^2_{\rm d}}.
\end{equation}
Thus, $E_{\rm c}$ is a local quantity which depends on the length of the dot
and the interaction strength in the dot region through $g_{\rm d}$. Likewise,
the level spacing $\epsilon=2 g_{\rm d}E_{\rm c}$ is a local
probe of the interaction that depends only on $g_{\rm d}$.

In order to calculate the conductance (\ref{cbcond}), the entire
spectral density, i.e., the Green's function at arbitrary frequency, is
needed. Treating $\varphi(x)$ perturbatively (up to first order) in
(\ref{eom}) we get
\begin{equation}
G(\omega;x_{\rm d}, x_{\rm d}) = \frac{i}{2}\frac{g_0}{\omega} \left( 1 + c \, \int_0^\infty {\rm d}y \left[ \varphi(y+x_{\rm d}) +\varphi(x_{\rm d}-y) \right] e^{i \eta y} \right),
\label{pgreens}
\end{equation}
where $\eta=2g_0 \omega/v_{\rm F}$ and $c =i2g_0^3\omega/2\pi v_{\rm
  F}^2$.  Equation (\ref{jdot}) remains valid for a small dot in an
inhomogeneous wire \cite{prl}, but $J_{\rm d}(\omega)$ is now deduced
from (\ref{pgreens}).  The rates entering expression (\ref{cbcond})
for the conductance must be computed numerically in the inhomogeneous
case.
Figure \ref{edin1} shows the temperature dependence of the CB peak
maximum for $\varphi(x)=[1+(2x/L^*)^2]^{-1}$, $x_{\rm d}=$ 0, $g_0=$
0.6 and $g_{\rm d}=$ 0.3. We observer a crossover from a power-law
with exponent $1/g_{\rm d}-2$ at high temperatures to one with
exponent $1/g_0-2$ at low temperature.  The measurement of the linear
conductance at low temperatures thus reflects the interaction far away
from the dot ({\em global probe}).  Measurements at high temperature,
on the other hand, act as a {\em local probe} of the interaction close
to the dot. It is to be noted that the transition region around
$T=\omega^\ast$ appears to cover at least one order of magnitude.
Fitting to a power-law in a narrow interval within the transition
region may yield any value $g_{\rm eff}$ between $g_{\rm d}$ and
$g_0$, related to the function ${\cal G}^{\rm max}_{\rm d}(T)$ through
$1/g_{\rm eff} = {\partial \ln {\cal G}^{\rm max}_{\rm d}/ \partial
  \ln T}+2$.
\section{Carbon nanotube SET}
In metallic nanotubes two one-dimensional bands intersect the Fermi
energy, hence two charge and spin channels are available for
transport. Only the symmetric combination of the two charge modes is
affected by the interaction \cite{egger}, characterized by the
interaction parameter $g_{\rm nt}\approx 0.27$ \cite{bockrath397}.
Taking into account the three non-interacting channels (due to the
spin modes and the antisymmetric combination of the charge modes), one
can account for this particular band structure by substituting
\cite{egger} $1/g_0\rightarrow(3+ 1/g_{\rm nt})/4$ in the above result
(\ref{dotcond}) for the dot-conductance ${\cal G_{\rm d}}$.

\begin{figure}[bt]
\begin{center}
\includegraphics[width=14cm,keepaspectratio]{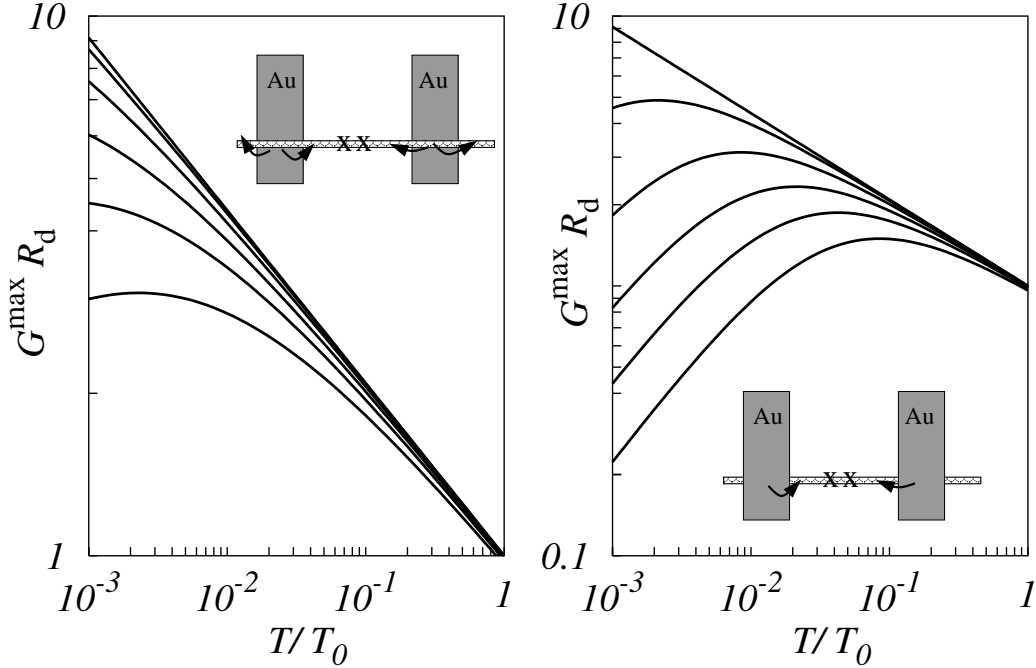} 
\end{center}
  \caption{
    Left: "bulk-tunneling" from Au-contacts into a nanotube with two
    impurities (tube on top of contacts). Temperature dependence of a
    conductance peak for a one-dimensional quantum dot in a TLL
    connected to leads through resistive tunnel contacts with
    resistances $R_{\rm d}$ (dot) and $R_{\rm c}$ (contacts) in units
    $R_{\rm d}$, $\alpha_{\rm bulk}=$0.25 ($g_{\rm nt}=$0.27). The
    curves correspond to ratios $R_{\rm c}/R_{\rm d}=$0, 0.0005,
    0.002, 0.005, 0.01, 0.02 (from top to bottom). Right:
    "end-tunneling" from Au lead into nanotube with quantum dot
    (contacts deposited on top of tube as in the room temperature SET
    setup). CB peak maximum for the same parameters as in left but the
    tunneling from the contacts into the nanotube is assumed to be
    into the end of the tube with the corresponding $\alpha_{\rm
      end}=$0.68. The low temperature power-law stems from the
    resistive tunnel contacts while the high-temperature power-law is
    due to the quantum dot.  }
\label{nano}
\end{figure}

Unlike CEO wires, nanotubes are very difficult to connect to leads
\cite{bockrath397}; the connection between the nanotube and a metallic
contact is usually regarded as a tunnel
junction\cite{kane,egger,matveev}. In the experiments done so far, two
kinds of contacts have been used. On the one hand, a contact can be
created by depositing metallic leads on top of the tube
(figure \ref{nano}, right sketch). In this case, the one-dimensional
conductor terminates at the contact, and tunneling occurs from the
external lead into the end of the nanotube. On the other hand,
nanotubes can be placed on top of predefined metallic leads. Then
electrons can tunnel from the metal into the one-dimensional bulk of a
tube (figure \ref{nano}, left). In both cases the temperature
dependence of the contact resistance is given by a TLL power-law of
the form $T^\alpha$ \cite{kane,egger,matveev}.  Tunneling into the
bulk of the nanotube is described by the exponent $\alpha_{\rm
  bulk}=(1/g_{\rm nt} +g_{\rm nt}-2)/8$ while tunneling into the end
corresponds to an exponent $\alpha_{\rm end}=(1/g_{\rm nt}-1)/4$. 
\pagebreak

We write the contribution of the contacts through tunneling into the
nanotube as the conductance
\begin{equation}
  \label{contact}
{\cal G}_{\rm c}^{\rm nt}(T)=\frac{1}{R_{\rm c}}\,
\left(\frac{T}{T_0}\right)^\alpha,
\end{equation}
where is $R_{\rm c}$ the resistance of the junction at an arbitrary reference
temperature $T_0$ (e.g., room temperature), and $\alpha$ is either the end or
bulk exponent.

The temperature dependence of the CB peak conductance is governed by a
different power-law,
\begin{equation}
  \label{dotntmax}
{\cal G}^{\rm max}_{\rm d}(T)=\frac{1}{R_{\rm
  d}}\left(\frac{T}{T_0}\right)^{\alpha_{\rm d}}. \nonumber
\end{equation}
where $\alpha_{\rm d} = (1/g_{\rm nt} - 5)/4$. Note that $\alpha_{\rm
d}<0$ for $g_{\rm nt} > 1/5$, while $\alpha_{\rm bulk}$ and
$\alpha_{\rm end}$ are always positive for repulsive interaction.

Assuming the electronic coherence length to be short compared to the extension
of the nanotube, the total resistance is given by the sum of the resistances
of the contacts and the dot,
\begin{equation}
{1\over {\cal G}(T,\mu)} = {2\over {\cal G}_{\rm c}^{\rm nt}(T)}
+ {1\over {\cal G}_{\rm d}(T,\mu)}.
\end{equation}
At room temperature the experimentally observed resistance $R_{\rm d}$ of
the dot is one order of magnitude larger than the combined contact
resistance $2 R_{\rm d}$ \cite{postma2000}. The high-temperature behaviour of
the observed conductance peaks is given by the exponent $\alpha_{\rm
d}$, and the transport properties of the wire are dominated by the SET
physics of the dot for arbitrary gate voltage.

At low temperatures, however, the resistance of the contacts dominates
over the relatively small resistance of the dot near a conductance
peak. This means that a two-point conductance measurement observes a
`clipped' peak, whose peak height is no longer governed by
(\ref{dotntmax}) but by (\ref{contact}).  Between the CB peaks the
resistance of the dot of the dot is always higher than that of the
contacts, with measured currents limited mostly by the Coulomb
blockade. Thus the temperature behaviour of the CB conductance {\em maxima}
can be given by the power-law of the {\em contact} conductance
(\ref{contact}) in spite of the fact that CB conductance peaks are
observed as a function of the gate voltage.

This implies a crossover between the two power-laws in the temperature
dependence of the CB peak, with low-temperature exponents,
$\alpha_{\rm end}$ or $\alpha_{\rm bulk}$, and crossover temperatures
\begin{equation}
    \label{eq:Tstar}
{T^*\over T_0} = 
\left( {2 R_{\rm c}\over R_{\rm d}} \right)^{1/(\alpha - \alpha_{\rm d})}
\end{equation}
where $\alpha$ is $\alpha_{\rm end}$ or $\alpha_{\rm bulk}$ depending on the
type of contact used in the experiment. For given parameters $R_{\rm c}$ and
$R_{\rm d}$ at the reference temperature $T_0$, the crossover temperatures are
related to each other by
\begin{equation}
  \label{eq:co_relation}
\left( T^*\over T_0 \right)_{\rm bulk}
 = \left( T^*\over T_0 \right)_{\rm end}
^{1/(\alpha_{\rm bulk} - \alpha_{\rm d})}.
\end{equation}
For the experimental value $g_{\rm nt}= 0.27$ and $R_{\rm c} \ll R_{\rm d}$
this means $T^*_{\rm bulk} \ll T^*_{\rm end}$. The behaviour of the conductance
maximum as a function of temperature is shown in figure ~\ref{nano}.

With the above model, the temperature behaviour of the Coulomb peak
described in \cite{postma2001} can be understood using the interaction
parameter $g_{\rm nt}=0.27$ reported earlier. Assuming metal-to-end
tunneling at the contacts, we find a quantitative agreement with the
observed increase of the peak conductance with rising temperature. The
authors of \cite{postma2001} discuss an alternative explanation of
their data, postulating a correlated tunneling process depending only
on intrinsic properties of the quantum dot. However, in view of the
existing knowledge about the behaviour of metal-to-nanotube contacts,
we consider the straightforward explanation given here rather
persuasive.
\section{Conclusion}
We demonstrate that contacts to quantum wires strongly influence
transport through a one-dimensional quantum dot system. Recent
experiments on both cleaved-edge overgrowth and carbon nanotube
quantum wires can thus be reconciled with the Tomonaga-Luttinger Model
of the one-dimensional electron liquid.  Non-Fermi-liquid transport
properties such as power-laws governing the height of Coulomb blockade
peaks were previously interpreted only through characteristics of the
quantum dot itself.  However, the power-laws in the temperature
dependence of the CB maxima can be modified or superseded by the
effect of inhomogeneous contacts (CEO's) or tunnel junctions
(nanotubes).  Depending on the energy regimes where the measurements
are carried out, the linear conductance is either a local probe
establishing the interaction parameter near the dot, or a global probe
of a larger system including extended one-dimensional excitations and
external contacts. The temperature distinguishing between these two
regimes is either determined by the length scale on which an
inhomogeneous interaction changes (CEO's), or by the ratio of the
contact and buckle tunneling resistances (nanotubes).


\section*{References}
\begin{thebibliography}{9}
\bibitem{aus} Auslaender O M et al 2000 Phys.\ Rev.\ Lett. 84 1764--8
\bibitem{kleim} Kleimann T et al 2000  Phys.\ Rev.\ B 62 8144--53
\bibitem{postma2001} Postma H W Ch et al 2001 Science 293 76--9
\bibitem{tans} Tans S J et al 1997 Nature 386 474--7
\bibitem{postma2000} Postma H W Ch et al 2000 Phys.\ Rev.\ B 62 R10653--6
\bibitem{bockrath291} Bockrath M et al 2001 Science 291 283--5
\bibitem{voit} Voit J, 1995 Rep.\ Prog.\ Phys.\ 58 977-1116
\bibitem{kane} Kane C L and Fisher M P A 1992 Phys.\ Rev.\ B  46 15233--62
\bibitem{weiss} Weiss U 1999 Quantum Dissipative Systems (Singapore: World Scientific)
\bibitem{alepl} Braggio A et al 2000 Eur.\ Phys.\ Lett. 50 236--42
\bibitem{grade}Ingold G L and Nazarov Y 1992 Single Charge
    Tunneling (New York: Plenum Press) p.\ 21--106 
\bibitem{prl} Kleimann T et al unpublished
\bibitem{egger} Egger R and Gogolin A O 1998 Eur.\ Phys.\ J.\ B 3 281--300
\bibitem{bockrath397} Bockrath M et al 1999 Nature 397 598--601
\bibitem{matveev} Matveev K A and Glazman L I 1993  Phys.\ Rev.\ Lett.\ 70 990--3
\end{thebibliography}
\end{document}